\documentclass[12pt,preprint]{aastex}

\usepackage[]{natbib}

\newcommand{\rxte}{{\it RXTE}}

\newcommand{\tfe}{1E~1048.1--5937} 
\newcommand{\tfn}{1E~2259.1+586}

\def\lapp{\ifmmode\stackrel{<}{_{\sim}}\else$\stackrel{<}{_{\sim}}$\fi}
\def\gapp{\ifmmode\stackrel{>}{_{\sim}}\else$\stackrel{>}{_{\sim}}$\fi}

\begin{document}

\title{Anomalous X-ray Pulsar 1E~1048.1--5937: Pulsed Flux Flares and Large Torque Variations}

\author{Fotis P. Gavriil\altaffilmark{1} and Victoria M. Kaspi\altaffilmark{1,2} }

\altaffiltext{1}{Department of Physics,
McGill University,  Rutherford Physics Building, 3600 University Street, Montreal, QC, H3A 2T8, Canada.}
\altaffiltext{2}{Department of Physics and Center for Space Research,
Massachusetts Institute of Technology, Cambridge, MA 02139}

\begin{abstract}
We report on continued monitoring of the Anomalous X-ray pulsar (AXP)
\tfe\ using the \textit{Rossi X-ray Timing Explorer}.  We confirm that
this pulsar has exhibited significant pulsed flux variability.  The
principal features of this variability are two pulsed X-ray
flares.  Both flares lasted several months and had well-resolved
few-week--long rises. 
The long rise times of the flares are a 
 phenomenon not previously reported for this class of object.
The epochs of the flare peaks were MJD $52,218.8 \pm 4.5$ and $52,444.4
\pm 7.0$.  Both flares had shorter rise than fall times.  The flares 
had peak fluxes of  $2.21 \pm 0.16$ and $3.00 \pm 0.13$ times
the quiescent value.  We estimate a total
2--10~keV energy release of  $\sim 2.7\times 10^{40}$~ergs and $\sim
2.8\times 10^{41}$ ergs for the flares, assuming a distance of 5 kpc.  
We also report large (factor of $\sim$12) changes
to the pulsar's spin-down rate on time scales of weeks to months, shorter
than has been reported previously.  We find marginal evidence for
correlation between the flux and spin-down rate variability, with
probability of nonrandom correlation 6\%.  We discuss the implications
of our findings for AXP models.
\end{abstract}


\section{Introduction}
\label{sec:intro}

Anomalous X-ray pulsars (AXPs) are an exotic manifestation of young
neutron stars.  AXPs are known for their steady, soft X-ray pulsations in
the period range of 6--12~s.  The detection of X-ray bursts from
two AXPs has confirmed the common nature of these objects with that of
soft gamma repeaters \citep{SGRs; gkw02,kgw+03}, another exotic type of young
neutron star.  Both classes of objects are believed to be magnetars,
i.e., powered by the decay of an ultrahigh magnetic field that has
a magnitude of $10^{14}$--$10^{15}$ G on the stellar surface.  For recent
AXP reviews, see \citet{kg04a} and \citet{kas04}.

One issue in AXP research has been flux stability.  Historically,
two AXPs have been reported to be highly flux variable.  \citet{opmi98}
collected all published flux measurements for AXP \tfe\ and concluded
that its total flux varies by as much as a factor of 10 between observations
spaced by typically 1--2 yr over $\sim$20 yr.  Those data were from a
diverse set of instruments, including imaging and nonimaging
telescopes.  Similarly, flux variability by a factor of greater than 4 was
reported for AXP \tfn\ by \citet{bs96}, using data also from a variety
of instruments.

However, long-term {\it Rossi X-Ray Timing Explorer (RXTE)} monitoring 
of the pulsed flux  of \tfe\ by
\citet{kgc+01} and of \tfn\ by \citet{gk02} using a single instrument
and set of analysis software showed no evidence to support such large
variability.\footnote{Total flux measurements with \rxte\ were difficult 
given the large field of view of the PCA and  the low 
count rates for the AXPs relative to the background.} 
Also, \citet{tgsm02}, following a short {\it
XMM-Newton} observation of \tfe, compared the observed flux with those
measured by two other imaging instruments, {\it ASCA} and {\it
BeppoSAX}.  They found that in the three observations, the total flux
was steady to within $\sim$30\%--50\%.  They argued that the nonimaging
detections included in the \citet{opmi98} analysis may have been
contaminated by other sources in the instruments' fields of view; in
particular, the bright and variable X-ray source $\eta$ Carina lies only
38$'$ away.

A possible solution to this puzzle came with the discovery of a large
(greater than 10 times) long-lived flux enhancement from \tfn\ at the time of a
major outburst in 2002 June 18. This event was accompanied by many other
radiative changes as well as by a large rotational spin-up
\citep{kgw+03,wkt+04}.  This suggests that past flux variability
reported in AXPs could be attributed to similar outbursts that went
undetected.

We report here, using data from our continuing {\it RXTE} monitoring
program, the discovery of significant pulsed flux variability in
\tfe.  This variability is mainly characterized by two long-lived pulsed 
flux flares, having well-resolved rises a few weeks long. These are unlike
any previously seen flux enhancements in AXPs and SGRs and thus likely
represent a distinct physical phenomenon.  We find no evidence for any
major associated bursting behavior.  We also report large variations
in the spin-down torque on timescales of a few weeks/months.  We find only
a marginal correlation between the flux and torque variations.  We
argue that this poses another significant challenge to any 
disk-accretion model for AXPs, but is not inconsistent with the
magnetar model.

\section{Analysis and Results}
\label{sec:obs}

All observations reported here were obtained with the Proportional
Counter Array \citep[PCA][]{jsss+96} on board  \rxte.  The timing
observations described below are a continuation of those reported by
\citet{kgc+01}.  We refer the reader to that paper for details of the
analysis procedure.  This \rxte\ monitoring program has shown that,
in general, AXPs have sufficient stability for phase coherent timing
\citep[see][for a review]{kg04a}. \tfe\ is an exception.  For this
pulsar, we have achieved phase-coherent timing only over relatively
short data spans.  In 2002 March, we adopted the strategy of
observing this source every week with three short ($\sim$2~ks)
observations.  These closely spaced observations allow us to measure
the spin frequency with high precision weekly without phase connecting
over long baselines.  This therefore allows us to determine the
spin-down rate with interesting precision on timescales of a few weeks.
Figure~\ref{fig:spin flux spectra}A shows the long-term spin history of
\tfe\ as measured by {\it RXTE}.

Figure~\ref{fig:fit}A shows the spin-down rate $\dot{\nu}$
as a function of time over the interval for which we can make this
measurement.  Plotted values of $\dot{\nu}$ were calculated by
measuring the slopes of each 5 adjacent values of
${\nu}$.  Note how $\dot{\nu}$ clearly varies greatly during our
observations, on all timescales to which we are sensitive. 
From MJD 52,400 to MJD 52,620 $\dot{\nu}$ had changed by a factor of $\sim$12.
During the
$\sim$120 day interval from MJD 52,620 through MJD 52,740, $\dot{\nu}$ was a
factor of $\sim$4 larger than the long-term average spin-down 
($\langle \dot{\nu} \rangle=-6.48\times 10^{-13}\ \mathrm{Hz}\ \mathrm{s}^{-1}$).  This
was followed by an abrupt decrease in magnitude by a factor of $\sim$2,
which was not resolved, and by subsequent additional variations.  
At no time did we observe any episode of spin-up.

We also monitor the pulsed flux of this source.  In this analysis, data
from each observing epoch were also folded at the optimal pulse
period.
We calculated the rms pulsed flux using the method described 
by \citet{wkt+04}.\footnote{In eq. (1) of \citet{wkt+04} there is  a typographical error of a  factor of 2 missing from the coefficients $\alpha_k$ and $\beta_k$;  similarly a factor of 4 is missing from their respective variances. The calculations in that paper however, used the correct form of the equation.}
Given \tfe's highly sinusoidal pulse profile we used  only 
the first two harmonics to calculate the pulsed flux.
This method of
measuring flux is different from the one used in  \citet{kgc+01} which
involved fitting a spectral model to extract a pulsed flux in cgs
units.  Given the  short length of the observations,  fitting a spectral
model to the individual observations was not practical.
The pointing for two observations was slightly off-source 
so we had to correct for reduced collimator response. 
The pointing was on-source for all other observations.
Figure~\ref{fig:spin flux spectra}B shows our pulsed flux time series
in the 2--10~keV band.  Pulsed flux time series in the 2--4 and
4--6~keV bands look similar.

The pulsed flux time series clearly has significant structure.  The
most obvious features are two long-lived flares.  The first flare was
smaller and shorter lived than the second.   The latter clearly
displayed significant structure in its decay.  In estimating the following flare
properties, we define the first flare as having occurred between MJDs
52,198 and 52,318 and the second having started on MJD 52,386, and we
take its end to be our last observation on MJD 53,030, although it
clearly has not yet ended (see Fig.~\ref{fig:fit}).  We estimate that
the first flare had a peak flux of  $2.21\pm 0.16$ times the quiescent pulse flux,
with the peak occurring at MJD $52,218.8 \pm 4.5$.  Its rise time was $20.8
\pm 4.5$ days, and its fall time $98.9 \pm 4.5$ days.  The second flare
peak was on MJD 52,444.4$\pm$7.0, and had a peak value of $3.00 \pm 0.13$
times the quiescent pulsed flux.  Its rise time was $58.3 \pm
7.0$~days, and its fall time is greater than 586 days.  We estimate 2--10~keV 
fluences of $(111 \pm 12)\times
10^4\ \mathrm{counts}\ \mathrm{PCU}^{-1}$ and  $(1136 \pm 38)\times
10^4\ \mathrm{counts}\ \mathrm{PCU}^{-1}$  for the first and second flare,
respectively.  \citet{tgsm02}  measured a total flux in the
2--10~keV energy range of $\sim 5 \times 10^{-12}$ ergs cm$^{-2}$~s$^{-1}$ and a
pulsed fraction of $\sim94\%$ (for energies $>$2 keV) from
\textit{XMM-Newton} observations of \tfe. This information, along with
our measured quiescent pulsed flux, allows us to scale our fluences to
estimate the total energy released in each flare. Assuming a distance
of 5 kpc \citep[see discussion in ][]{opk01}, we find a total energy
release of $\sim 2.7 \times 10^{40}$ ergs for the first flare and $\sim
2.8\times 10^{41}$ ergs for the second flare, both in the 2--10 keV
band.

Although we clearly detect both large flux variations and large changes
in the spin-down rate, the correlation between the two is marginal.
The Spearman rank order correlation coefficient $r_S = 0.28$, where 0
indicates no correlation and 1 indicates total correlation.  The
probability of obtaining this value of $r_S$ or higher by random chance
is 6\%.  Thus, there is marginal evidence of some correlation,
equivalent to a $\gapp 2\sigma$ result.  From Figure~2, it is clear why
any correlation is not strong:  for example, $\dot{\nu}$ changes very
little during the rise of the second flare, in the interval MJD
52,380-52,420.  Also, there is no short-term flux change when
$\dot{\nu}$ suddenly reaches its maximum absolute value (near MJD
52,620), nor when it abruptly changes by a factor of $\sim$2 around MJD
52,740.

Hardness ratios (HRs) were measured by comparing the pulsed flux, as
measured by the method described above, in the 2--4 keV band to that in
the 4--6 keV band.  Figure~\ref{fig:spin flux spectra}C shows our HR
measurements.  The mean HR is 0.78.  There is evidence for
spectral variability.  The reduced $\chi^2$ of the HR time series is
3.6 for 143 degrees of freedom.  However, there is no evidence for any
correlation of HR with pulsed flux or torque.  Our uncertainties,
however, are quite large; monitoring observations with an imaging
instrument would improve this situation.

Intriguingly, the peak of the first flare was coincident with the
epochs during which we observed two SGR-like X-ray bursts from the
direction of this source in 2001
\citep[][indicated by arrows in Fig.~\ref{fig:spin flux spectra}]{gkw02}.  
However, we found no other
SGR-like 
bursts in any of the remaining data.  For a detailed description of our
burst-searching algorithm see \citet{gkw04}.  We also searched our
folded time series for pulse morphology variations using the method
detailed by \citet{gk02}.  We find no evidence for significant pulse
profile changes at any epoch in our data set.

\section{Discussion}
\label{sec:discussion}

The long-lived flux enhancements with well-resolved rises that we have
observed in \tfe\ are very different from previously detected X-ray
flux variations in AXPs and SGRs, which show very abrupt rises
associated with major outbursts \citep[e.g.][]{kgw+03,wkt+04}.  
The long-lived flux decay in those
sources has been attributed to burst afterglow, which is a cooling of the
crust following an impulsive heat injection from magnetospheric bursts
\citep{let02}.  The much more gradual flux rises that we have observed in
\tfe\ comprise a new phenomenon not yet observed in any other AXP,
despite several years of careful and frequent {\it RXTE} monitoring.
These flux variations may provide a new diagnostic of the physical
origin of the persistent nonthermal emission in SGRs and AXPs, since
they are not contaminated by burst afterglow.
Also interesting are the large variations in spin-down rate
or torque.  Torque variations by nearly a factor of 5 were already 
reported from {\it RXTE} observations \citep{kgc+01}, on timescales of years.
Here we have shown that the torque can change
by at least a factor of $\sim$2 more, and on much shorter timescales,
namely, a  few weeks to months.

\begin{figure}
\plotone{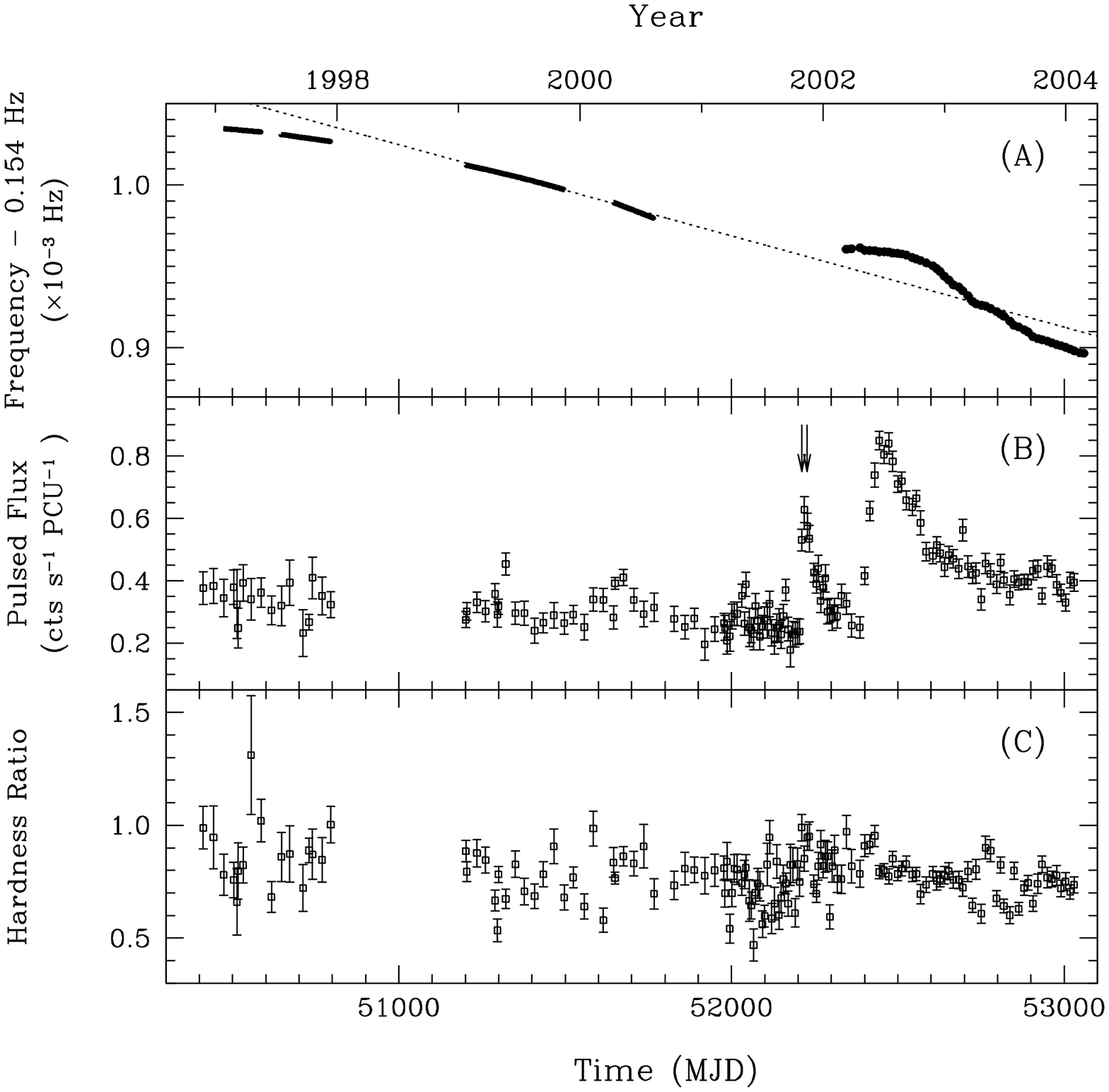}
\figcaption{Spin, flux, and spectral history of \tfe.  \textit{(a)} Observed spin
frequencies vs. time. The points represent individual frequency
measurements. The solid lines represent the phase-connected intervals
as reported by \citet{kgc+01}.  The dashed line is the long-term average
spin-down.  \textit{(b)} Pulsed flux time series in the 2--10~keV
band. Arrows indicate the times at which  the bursts reported by
\citet{gkw02} occurred. \textit{(c)} HR as a function of time. The
HRs displayed were computed for the pulsed flux in the energy range
(4--6~keV)/(2--4~keV).
\label{fig:spin flux spectra}}
\end{figure}

\begin{figure}
\plotone{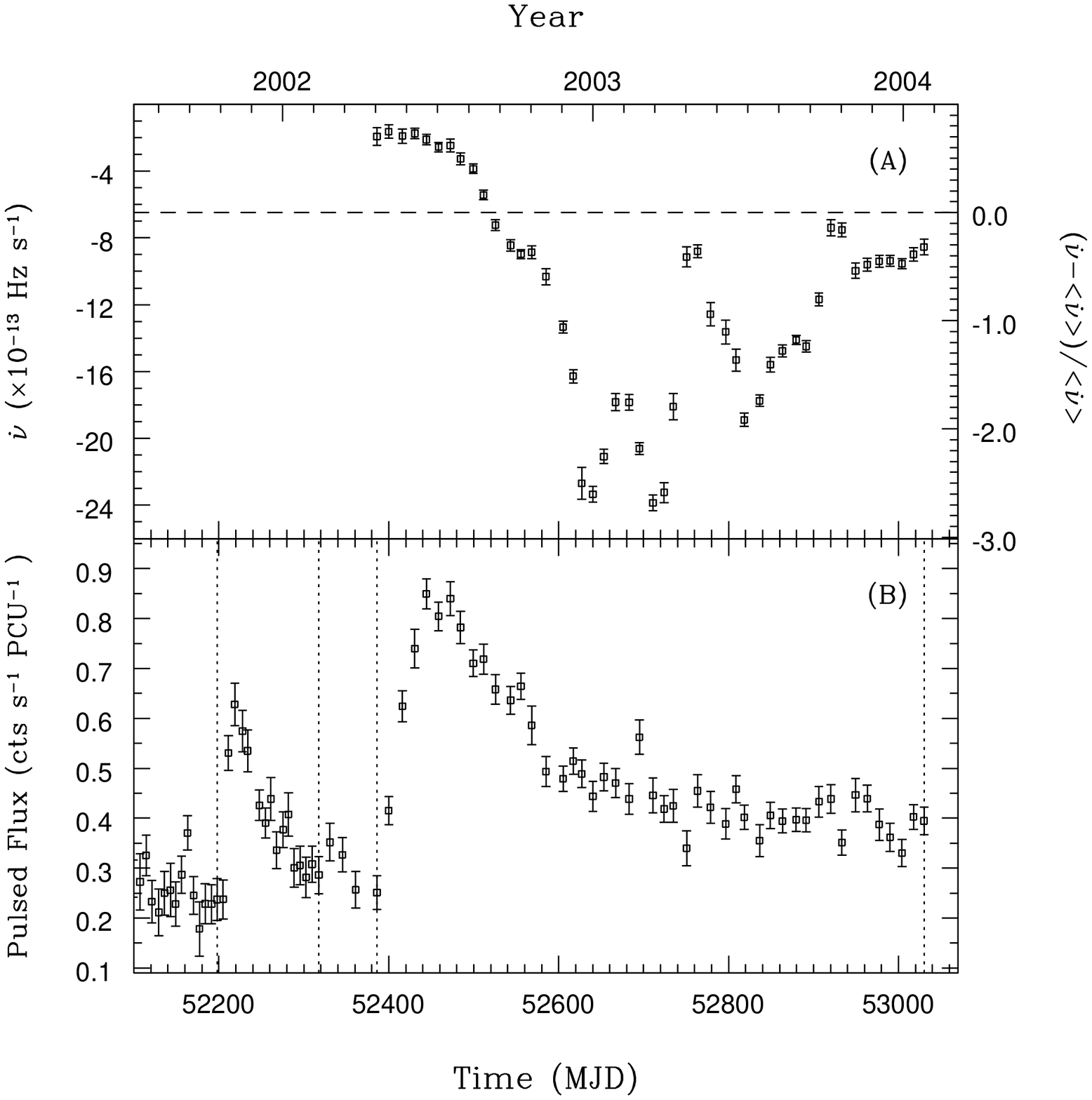}
\figcaption{\textit{(a)} $\dot{\nu}$ vs.  time over the interval for which
our data allow the measurement.  The horizontal dashed line denotes the
long-term average spin-down rate, $\langle \dot{\nu} \rangle$.  The right-hand scale is 
the fractional  difference  of $\dot{\nu}$ and the long-term average spin-down rate.
\textit{(b)} Zoom-in of the pulsed flux time series in the 2--10~keV band. 
Vertical dotted lines denote the chosen start and end ranges for characterizing
the two principal flares.
\label{fig:fit}}
\end{figure}

In considering the observed pulsed flux and torque variations, whether
they are correlated is an important issue.  Our weekly monitoring of
the source unfortunately commenced only after most of the first flare
decayed.  Prior to that, the monthly observations, taken in the form of
brief snapshots, did not allow anything about the rotational behavior
of the source to be determined when phase-coherent timing was not
possible.  This was the case during the first flare.  During the second
flare, the spin frequency was, interestingly, {\it most} stable during
the rise and peak of the flare.  Furthermore, the stable spin-down
rate was at a lower magnitude than the long-term average.
Subsequently, $\sim 60$ days after the flux began to decay, the rate of
spin-down began to increase.  Given timing observations during only one
flare, it is unclear whether these features are coincidences or not.  However,
there is no strong evidence to support otherwise; similar torque
variations were seen in the past and were not accompanied by any
flaring (see Fig.~\ref{fig:spin flux spectra}). 
Significant torque variations  unaccompanied by severe 
flux variability have been noted for \tfe\  prior to our \rxte\ monitoring
\citep[e.g.][]{pkdn00}. 
Nevertheless, statistically, the probability that
they are uncorrelated is only 4\%; studying Figure~\ref{fig:fit}
suggests that if anything, slope transitions are correlated, if not the
slopes between transitions.  Continued \rxte\ monitoring will help
identify any true correlations, particularly if the source exhibits
more variability.

Can the magnetar model explain such behavior?  The persistent emission in
magnetars has a spectrum that is well described by a two-component
model, consisting of a blackbody plus a hard power-law tail.
The thermal component is thought to arise from heat resulting from the
active decay of a high internal magnetic field \citep{td96a}; however,
thermal X-ray flux changes are not expected on as short a time scale as
we have measured in the absence of major bursts.  
\citet{tlk02} put forth a model in which the
nonthermal component arises from resonant Compton scattering of
thermal photons by currents in the magnetosphere.  In magnetars, these
currents are maintained by magnetic stresses acting deep inside its
highly  conducting interior, where it is assumed that the magnetic field
lines are  highly twisted.  These magnetospheric currents in turn twist
the external dipolar field in the lesser conducting magnetosphere.
These magnetic stresses can lead to sudden outbursts or more gradual
plastic deformations of the rigid crust, thereby twisting the
footpoints of the external magnetic field and inducing X-ray luminosity
changes.  The persistent  non-thermal emission  of AXPs is explained in
this  model  as being  generated by these currents through
magnetospheric Comptonization and surface back-heating
\citep{td96a,tlk02}.  Changes in X-ray luminosity, spectral hardness,
and torque have a common physical origin in this model and some
correlations are expected.  Larger twists correspond to harder
persistent X-ray spectra, as is observed, at least when comparing the
harder SGR spectra to those of the softer AXPs.  As noted by
\citet{kgc+01}, \tfe's hard photon spectral index ($\Gamma = 2.9$)
suggests that it is a transition object between the AXPs ($\Gamma \simeq$3--4) 
and the SGRs ($\Gamma = $2.2--2.4). Hence, if during the flares
\tfe's magnetosphere was twisted to the SGR regime, we
expect spectral index variations of $\sim 0.5$.  Spectral measurements
of such precision are not feasible with our short \rxte\ monitoring
observations.  

Decoupling between the torque and the luminosity can be
accounted for in the magnetar model.  According to \citet{tlk02}  the
torque is most sensitive to the current flowing on a relatively narrow
bundle of field lines that are anchored close to the magnetic pole, and
so only a broad correlation in spin-down rate and X-ray luminosity
is predicted, and in fact is observed for the
combined population of SGRs and AXPs \citep{mw01,tlk02}.  However, for a
single source, whether an X-ray luminosity change will be accompanied
by a torque change depends on where in relation to the magnetic pole
the source of the enhanced X-rays sits.  Similarly, large torque variations,
as we have observed, may occur in the absence of luminosity changes
if the former are a result of changes in the currents flowing only
in the small polar cap region.  

Note that energetically, the total release in these flares is
comparable to, although somewhat less than, that in the afterglows seen in
SGRs and in AXP \tfn\ \citep[see][for a summary]{wkt+04}.  It 
easily can  be accounted for given the inferred magnetic energy of the
star.

Although the magnetar model for AXPs has been spectacularly successful
in explaining their most important phenomenology, the anomalous behavior noted
for \tfe\ raises the possibility that perhaps it has a physical nature
different from other AXPs.  It has also been suggested
that AXPs might be powered by accretion from fossil disks \citep{chn00,
alp01}.  An increase in luminosity $L_X$ can easily be explained in accretion
models by an increase in the mass accretion rate $\dot{M}$, given that
$L_X \propto \dot{M}$.  Transient changes in $\dot{M}$ are perhaps not
unreasonable to expect in fossil disk models, given the huge variations
seen in $\dot{M}$ of conventional accreting sources.  However, in an
accretion scenario, we expect correlations
between luminosity and torque.  In conventional disk-fed accreting pulsars
undergoing spin-up, one expects $\dot{\nu} \propto L_X^{6/7}$.
Such a correlation is seen approximately in
accreting pulsars, with discrepancies possibly attributable to changed
beaming or improper measurement of bolometric luminosities, the former
due to pulse profile changes, and the latter due to finite
bandpasses \citep{bcc+97}.  As discussed by \citet{kgc+01}, for a
source undergoing regular {\it spin-down} as in \tfe, the prediction is less
clear; the form of the correlation depends on the unknown functional
form of the torque.  For the propeller torque prescription of
\citet{chn00}, we find that $L_X\propto \dot{\nu}^{7/3}$, a much stronger
correlation than in the conventional spin-up sources.  For a change
in $L_X$ by a factor of $\sim$3 as we have seen in the rise of the
second flare, we would expect
a simultaneous change in $\dot{\nu}$ by greater than $50\%$, 
clearly ruled out by
our data.  Conversely, for the abrupt change of $\dot{\nu}$ by a factor of 
$\sim$2 (near MJD 52740), we expect a change in $L_X$ by a factor
of $\sim$5, definitely not seen.  This appears to
pose a significant challenge to fossil-disk accretion models
for \tfe.

Two infrared  observations taken  on MJD 52,324 \citep{ics+02}
and MJD 52372 \citep{wc02} 
have shown that the IR counterpart 
of this source is variable. However, the pulsed 
X-ray flux at both those epochs was consistent with the quiescent value.
Furthermore, even though the X-ray flux has not yet returned to its
quiescent value, recent observations show that the source's proposed IR
counterpart is consistent with the fainter of the two previous observations
\citep{dk04}.  This
decoupling  between the IR and the X-ray flux contrasts with what
was observed in AXP \tfn, whose IR flux increased then decayed in
concert with the X-ray flux at the time of its 2002 outburst
\citep[][Tam et al. 2004, in preparation]{kgw+03}.  This is
puzzling and suggestive of more than one mechanism for producing
IR emission in AXPs.

\acknowledgements
We thank M. Lyutikov, S. Ransom, M.~Roberts, C. Thompson and
P.~Woods for useful discussions.  This work is supported by NSERC
Discovery Grant 228738-03, NSERC Steacie Supplement 268264-03, a Canada
Foundation for Innovation New Opportunities Grant, FQRNT Team and
Centre Grants, and NASA Long-Term Space Astrophysics Grant NAG5-8063.
V.M.K. is a Canada Research Chair and Steacie Fellow.  This research
has made use of data obtained through the High Energy Astrophysics
Science Archive Research Center Online Service, provided by the
NASA/Goddard Space Flight Center.

\bibliographystyle{apj}

\begin{thebibliography}{23}
\expandafter\ifx\csname natexlab\endcsname\relax\def\natexlab#1{#1}\fi

\bibitem[{Alpar (2001)}]{alp01}
Alpar, M. A.  2001, ApJ, 554, 1245

\bibitem[{Baykal \& Swank(1996)}]{bs96}
Baykal, A. \& Swank, J. 1996, ApJ, 460, 470

\bibitem[{{Bildsten} {et~al.}(1997){Bildsten}, {Chakrabarty}, {Chiu}, {Finger},
  {Koh}, {Nelson}, {Prince}, {Rubin}, {Scott}, {Stollberg}, {Vaughan},
  {Wilson}, \& {Wilson}}]{bcc+97}
{Bildsten}, L., {Chakrabarty}, D., {Chiu}, J., {Finger}, M.~H., {Koh}, D.~T.,
  {Nelson}, R.~W., {Prince}, T.~A., {Rubin}, B.~C., {Scott}, D.~M.,
  {Stollberg}, M., {Vaughan}, B.~A., {Wilson}, C.~A., \& {Wilson}, R.~B. 1997,
  ApJS, 113, 367

\bibitem[{{Chatterjee} {et~al.}(2000){Chatterjee}, {Hernquist}, \&
  {Narayan}}]{chn00}
{Chatterjee}, P., {Hernquist}, L., \& {Narayan}, R. 2000, ApJ, 534, 373

\bibitem[{{Durant} \& {van Kerkwijk}(2004)}]{dk04}
{Durant}, M. \& {van Kerkwijk}, M.~H. 2004, in 
Young Neutron Stars and Their Environments,
IAU Symposium 218, eds. B. Gaensler \& F. Camilo,
(San Francisco:  Astronomical Society of the Pacific),
in press (astro-ph/0309801)

\bibitem[{{Gavriil} \& {Kaspi}(2002)}]{gk02}
{Gavriil}, F.~P. \& {Kaspi}, V.~M. 2002, ApJ, 567, 1067

\bibitem[{Gavriil {et~al.}(2002)Gavriil, Kaspi, \& Woods}]{gkw02}
Gavriil, F.~P., Kaspi, V.~M., \& Woods, P.~M. 2002, Nature, 419, 142

\bibitem[{Gavriil {et~al.}(2004)Gavriil, Kaspi, \& Woods}]{gkw04}
---. 2004, ApJ, in press; astro-ph/0310852


\bibitem[{{Israel} {et~al.}(2002){Israel}, {Covino}, {Stella}, {Campana},
  {Marconi}, {Mereghetti}, {Mignani}, {Negueruela}, {Oosterbroek}, {Parmar},
  {Burderi}, \& {Angelini}}]{ics+02}
{Israel}, G.~L., {Covino}, S., {Stella}, L., {Campana}, S., {Marconi}, G.,
  {Mereghetti}, S., {Mignani}, R., {Negueruela}, I., {Oosterbroek}, T.,
  {Parmar}, A.~N., {Burderi}, L., \& {Angelini}, L. 2002, ApJ, 580, L143

\bibitem[{Jahoda {et~al.}(1996)Jahoda, Swank, Stark, Strohmayer, Zhang, \&
  Morgan}]{jsss+96}
Jahoda, K., Swank, J., Stark, M., Strohmayer, T., Zhang, W., \& Morgan, E.
  1996, Proc. SPIE, 2808, 59

\bibitem[{Kaspi(2004)}]{kas04}
Kaspi, V.~M. 2004, in Young Neutron Stars and Their Environments, {IAU}
  Symposium 218, eds. B.~Gaensler \& F.~Camilo (San Francisco: Astronomical
  Society of the Pacific), in press (astro-ph/0402175)

\bibitem[{Kaspi \& Gavriil(2004)}]{kg04a}
Kaspi, V.~M. \& Gavriil, F.~P. 2004, in The Restless High-Energy Universe, ed.
  E.~van~den Heuvel, J.~in't Zand, \& R.~Wijers (Elsevier), in press;
  astro-ph/0402176

\bibitem[{Kaspi {et~al.}(2001)Kaspi, Gavriil, Chakrabarty, Lackey, \&
  Muno}]{kgc+01}
Kaspi, V.~M., Gavriil, F.~P., Chakrabarty, D., Lackey, J.~R., \& Muno, M.~P.
  2001, ApJ, 558, 253

\bibitem[{Kaspi {et~al.}(2003)Kaspi, Gavriil, Woods, Jensen, Roberts, \&
  Chakrabarty}]{kgw+03}
Kaspi, V.~M., Gavriil, F.~P., Woods, P.~M., Jensen, J.~B., Roberts, M. S.~E.,
  \& Chakrabarty, D. 2003, ApJ, 588, L93

\bibitem[{{Lyubarsky} {et~al.}(2002){Lyubarsky}, {Eichler}, \&
  {Thompson}}]{let02}
{Lyubarsky}, Y., {Eichler}, D., \& {Thompson}, C. 2002, ApJ, 580, L69

\bibitem[{{Marsden} \& {White}(2001)}]{mw01}
{Marsden}, D. \& {White}, N.~E. 2001, ApJ, 551, L155

\bibitem[{Oosterbroek {et~al.}(1998)Oosterbroek, Parmar, Mereghetti, \&
  Israel}]{opmi98}
Oosterbroek, T., Parmar, A.~N., Mereghetti, S., \& Israel, G.~L. 1998, A\&A,
  334, 925

\bibitem[{\"Ozel {et~al.}(2001){\"Ozel}, {Psaltis}, \& {Kaspi}}]{opk01}
\"Ozel, F., Psaltis, D. \& Kaspi, V. M. 2001, ApJ, 563, 255

\bibitem[{Paul {et~al.}}(2000)]{pkdn00}
{Paul}, B., {Kawasaki}, M., {Dotani}, T., \& {Nagase}, F. 2000, ApJ, 537, 319

\bibitem[{Thompson \& Duncan(1996)}]{td96a}
Thompson, C. \& Duncan, R.~C. 1996, ApJ, 473, 322

\bibitem[{Thompson {et~al.}(2002)Thompson, Lyutikov, \& Kulkarni}]{tlk02}
Thompson, C., Lyutikov, M., \& Kulkarni, S.~R. 2002, ApJ, 574, 332

\bibitem[{{Tiengo} {et~al.}(2002){Tiengo}, {G{\" o}hler}, {Staubert}, \&
  {Mereghetti}}]{tgsm02}
{Tiengo}, A., {G{\" o}hler}, E., {Staubert}, R., \& {Mereghetti}, S. 2002,
  A\&A, 383, 182


\bibitem[{{Wang} \& {Chakrabarty}(2002)}]{wc02}
{Wang}, Z. \& {Chakrabarty}, D. 2002, ApJ, 579, L33

\bibitem[{Woods {et~al.}(2004)Woods, Kaspi, Thompson, Gavriil, Chakrabarty,
  Marshall, Flanagan, Heyl, \& Hernquist}]{wkt+04}
Woods, P.~M., Kaspi, V.~M., Thompson, C., Gavriil, F.~P., Chakrabarty, D.,
  Marshall, H.~L., Flanagan, K., Heyl, J., \& Hernquist, L. 2004, ApJ, 605, 378

\end{thebibliography}

\end{document}